\documentstyle[epsfig,epsf]{article}

\newcommand{\ov}{\overline}

\begin{document}
\large
\begin{flushright}
DFUB 2000-6 \par
Bologna, April 2000
\end{flushright}

\begin{center}
{\Large {\bf Magnetic monopoles, nuclearites, Q-balls: a qualitative picture}}
\end{center}

\vskip .7 cm

\begin{center}
D. Bakari$^{1,2}$, S. Cecchini$^{1}$, H. Dekhissi$^{1,2}$, J. Derkaoui$^{1,2}$,
 G. Giacomelli$^{1}$, \par M. Giorgini$^{1}$,  G. Mandrioli$^{1}$, 
 A. Margiotta$^{1}$, M. Ouchrif$^{1,2}$, \par L. Patrizii$^{1}$, 
V. Popa$^{1,3}$, P. Serra$^{1}$ and M. Spurio$^{1}$
\end{center}

\begin{center}
{\it $^1$Dipartimento di Fisica dell'Universit\`a di Bologna and INFN, Sezione 
di Bologna, I-40127 Bologna, Italy \\
$^2$Faculty of Sciences, University Mohamed I, B.P. 524 Oujda, Morocco \\
$^3$Institute of Space Sciences, Bucharest R-76900, Romania}
\vskip .7 cm
{\bf Abstract}\par
\end{center}

We present qualitative pictures of the structures of 
magnetic monopoles (MMs),
 nuclearites (nuggets of strange quark matter, strangelets, surrounded 
by electrons) and Q-balls 
(supersymmetric states of squarks, sleptons and Higgs fields).  In 
particular we discuss the relation between their mass and 
size. MMs, nuclearites and Q-balls   
could be part of the cold Dark Matter (DM); we consider 
astrophysical limits on the flux of these particles in the cosmic radiation.

\vspace{5mm}

\section{Introduction}

Magnetic monopoles (MMs), nuclearites (nuggets of strange quark 
matter + electrons) and Q-balls (supersymmetric states of 
squarks, sleptons and Higgs fields) could be part of the cold
dark matter (DM) located in the halo of galaxies. They would have typical
velocities of $\beta c \sim 10^{-3}c$; galactic magnetic fields may accelerate 
magnetic monopoles to larger velocities. \par
In this note we show some qualitative pictures of the structures of (i) 
superheavy GUT magnetic monopoles, (ii)
 intermediate mass magnetic monopoles, (iii) nuclearites and (iv) 
Q-balls. We discuss in particular the relation between the mass and the size of
these particles. \par
Based on astrophysical considerations, we discuss phenomenological 
limits on the fluxes in the 
cosmic radiation, of magnetic monopoles, nuclearites and Q-balls.

\section{Magnetic monopoles}
\label{par:monopoles}
The concept of magnetic monopole may be traced back to the origin of
magnetism. In 1931 Dirac introduced the MMs in order  to explain the
quantization of the electric charge, obtaining the formula 
$eg=n \hbar c/2$, from which $g=ng_D=n \hbar c/2e=68.5e=3.29 \cdot 10^{-8}$ 
c.g.s. symmetric system \cite{Dirac}. A MM possessing 
also an electric charge
is called a dyon. A MM and an atomic nucleus may form a bound system with
both magnetic and electric charges: also this system is called a dyon. \par
 The energy losses  of MMs and of dyons in matter, in the Earth 
and in different detectors were 
calculated in ref. \cite{loss-mono}. An extensive bibliography of MMs
is given in ref. \cite{biblio}.

\subsection{GUT magnetic monopoles}
\label{par:GUT}
In the context of the Grand Unified Theory (GUT) of electroweak and 
strong interactions,  MMs with magnetic charges $g=ng_D$
are predicted to appear at the cosmic time of
$\sim 10^{-34}$~s, during the phase transition
corresponding to the spontaneous breaking of the Grand Unified group 
\cite{Kibble}. 
 They should have been produced as point defects or in extremely high energy
collisions of the type $e^+ e^- \to M \ov M$ \cite{Kibble}. One of the 
main problems 
with GUT MMs is the too large abundance predicted by the standard 
cosmology. Models with inflation at the end of the GUT epoch 
reduce dramatically their number and we would be left mainly with MMs
produced in very high energy collisions. \par

\begin{figure}[ht]
\begin{center}
\mbox{
        \epsfig{file=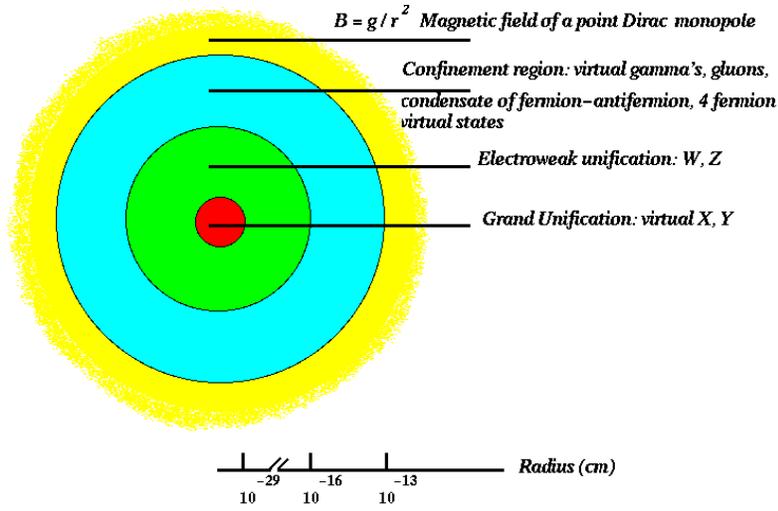,height=7cm}}
\end{center}
\vspace{-0.5cm}
\caption{Structure of a GUT monopole. The various regions correspond 
to: (i) Grand Unification ($r \sim 10^{-29}$ cm; inside this core one finds
virtual $X$ and $Y$ particles); (ii) electroweak unification 
($r \sim 10^{-16}$ cm; inside this region one finds virtual $W^{\pm}$ 
and $Z^0$); (iii)
confinement region ($r \sim 10^{-13}$ cm; inside one finds virtual 
$\gamma$, gluons and a condensate of fermion-antifermion pairs and 4-fermion
virtual states); (iv) for $r>$ few fm one has the field of a point 
magnetic charge.}
\label{fig:GUT}
\end{figure}

The GUT MM
mass is related to the mass of the X vector bosons, carriers of the 
unified interaction, by the relation $M_M \geq M_X/ \alpha$, where $\alpha
\simeq 0.025$ is the dimensionless unified coupling constant and $M_X \simeq
10^{14} \div 10^{16}$ GeV/c$^2$. Thus GUT magnetic monopoles should have 
$M_M \geq 10^{16}$ GeV/c$^2$. Due to
their large masses, these MMs cannot be produced with existing and 
foreseen accelerators. They must be searched for in the penetrating cosmic
radiation, using large area detectors \cite{Monopoli}. GUT MMs 
should be characterized by relatively 
low velocities and relatively large energy losses. Direct searches for GUT MMs
gave flux upper limits of few $10^{-16}$ 
cm$^{-2}$s$^{-1}$sr$^{-1}$ \cite{Monopoli}; several indirect
 limits were obtained by different experiments  \cite{biblio,GUT}. \par
The predicted spatial structure of a GUT MM is illustrated in Fig. 
\ref{fig:GUT} \cite{GUT}. The various regions are described in the 
figure caption. A GUT magnetic monopole may catalyse proton decay 
\cite{Kibble,GUT}; the cross
section for this process could be relatively large because of the size 
of the fermion-antifermion condensate, which may contain
 terms violating baryon number conservation (see Fig. \ref{fig:GUT}).

\subsection{Intermediate mass monopoles} 
\label{par:SLIM}
Magnetic monopoles with masses of $10^{10} \div 10^{12}$ GeV/c$^2$ are 
predicted
by theories with an intermediate mass scale \cite{scala}
and would appear in the early universe at a time considerably later than
the GUT time. Also these MMs are
 topological point defects; an undesirable large number of relatively light
monopoles may be gotten rid of by means of higher dimensional topological
defects (strings, walls, textures) \cite{scala}. \par

\begin{figure}[ht]
\begin{center}
\mbox{
        \epsfig{file=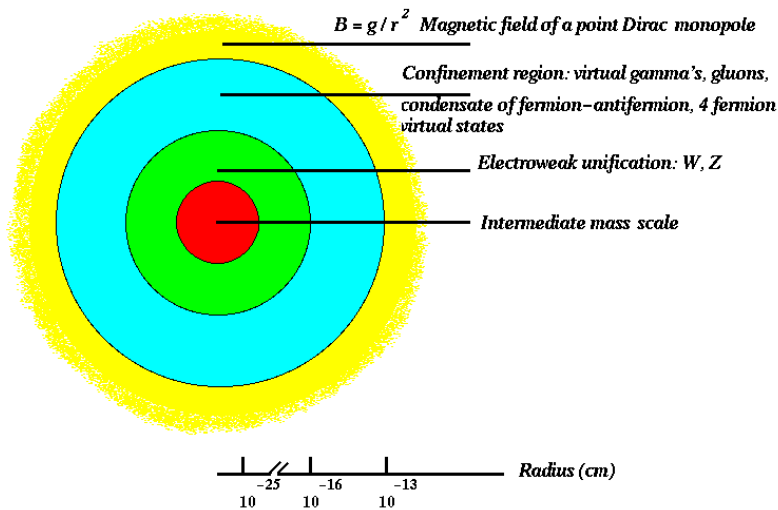,height=7cm}}
\end{center}
\vspace{-0.5cm}
\caption{Possible structure of an ``intermediate mass magnetic monopole". The 
inner region ($r \simeq 10^{-25}$ cm) corresponds to intermediate mass 
scales; inside this region one finds the intermediate mass bosons responsible
for the symmetry breaking. The outer regions are as in 
Fig. \ref{fig:GUT}, but without terms violating baryon number conservation 
in the fermion-antifermion condensate.}
\label{fig:SLIM}
\end{figure}

A possible structure of an intermediate mass  MM is illustrated in Fig. 
\ref{fig:SLIM}. Notice that it has a larger core compared to the structure 
of a GUT monopole. The various regions correspond to: (i) intermediate
 mass scale of $R \simeq 10^{-25}$ cm; (ii) the 
electroweak scale; (iii) the  condensate of
fermion-antifermion pairs which could be the same as for a GUT monopole, but 
without any
term violating baryon number conservation (thus these MMs cannot catalyse 
proton decay); (iv) the confinement region; (v) the outside region. \par
The number of intermediate mass monopoles 
could be considerably higher than that of GUT monopoles and galactic 
magnetic fields could accelerate 
them to high velocities. It has even be assumed that few of them could 
reach extremely high energies, interact in the upper earth
atmosphere and lead to the
highest energy cosmic ray showers \cite{weiler,escobar}.  

\section{Nuclearites}
\label{par:nuclearites}
 
Strange Quark Matter (SQM) should consist of aggregates 
of $u,~d$ and $s$ quarks in approximately 
equal proportions \cite{Witten}. The SQM is a colour singlet, thus
it may have 
only integer electric charges. The overall neutrality of SQM
is ensured by
an electron cloud which surrounds it, forming 
a sort of atom. SQM may be the ground state of QCD.

\begin{figure}[ht]
\begin{center}
\mbox{\epsfig{file=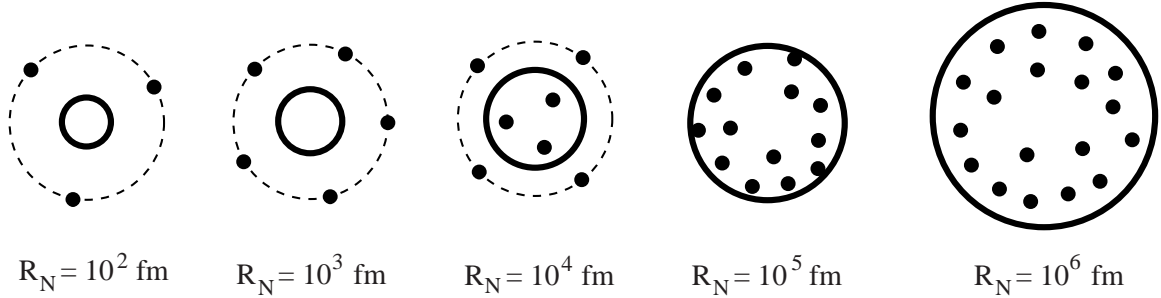,height=4cm}}
\end{center}
\vspace{-0.5cm}
\caption{Dimensions of the quark bag ($R_N$) and of 
the core+electrons system (nuclearite).
 The radii presented here (in a logarithmic scale) refer to the nuclearite
quark bag. For nuclearite 
masses smaller than $10^{9}$ GeV/c$^2$, the whole electron cloud is outside the
quark bag and the core+electrons system  has a global size
of approximately $10^{5}$ fm = 1 \AA; for $10^{9} < M_N < 10^{15}$ 
GeV/c$^2$ the electrons are partially inside the core; for $M_N > 10^{15}$ 
GeV/c$^2$ all electrons are inside the core. The black dots indicate the
electrons, the quark bag border is indicated by thick solid lines; the
border of the core+electronic cloud system for relatively small masses 
is indicated by the dashed lines.}
\label{fig:nucleariti}
\end{figure}

The aggregates of $u,~d,~s$ quarks will be
denoted with the terms  strangelet, quark bag, SQM and nuclearite
 core; we shall use the word nuclearite for the 
core+electrons system. \par
 Strangelets could have been produced shortly after the Big Bang and may
have survived as remnants; they could 
also appear in violent astrophysical
processes, such as neutron star collisions.
Nuclearites should have a constant density \cite{Rujula},
 $\rho_N = M_N/V_N \simeq 3.5 \cdot 10^{14}$ g cm$^{-3}$, somewhat 
larger than that of atomic nuclei.  
\par They should be stable for all baryon
numbers in the range between ordinary heavy nuclei and neutron stars 
($A \sim 10^{57}$) \cite{Rujula}. Nuclearites could contribute to the cold
dark matter.   \par

\begin{figure}
\vspace{-0.5cm}
\begin{center}
\mbox{  \epsfysize=8.5cm 
        \epsffile{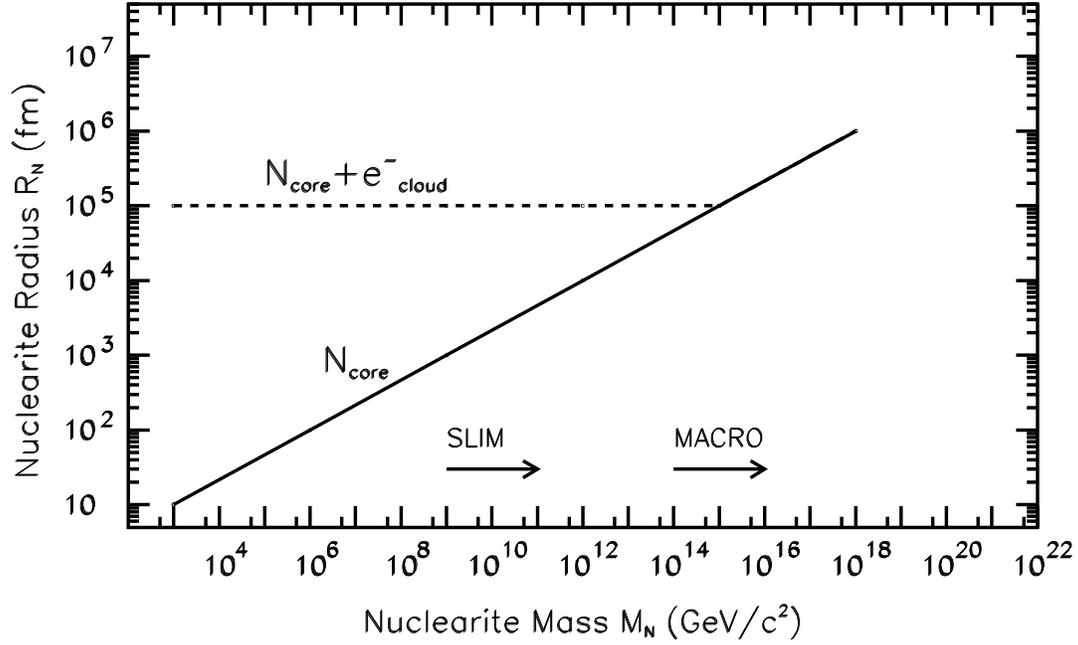}}
\end{center}
\vspace{-0.5cm}
\caption{Solid line: dependence of the nuclearite quark bag radius $R_N$ 
from its 
mass $M_N$. The dashed line gives the radius of the core+electrons 
system (nuclearite). We
also indicate the mass regions accessible by the MACRO
[15] and SLIM [23] experiments.}
\label{fig:r-vs-mass-nuc}
\end{figure}

The structure of SQM can be described in terms of a 
bag model \cite{Kasuya}. In order to equilibrate the chemical potential
of the quark species, SQM should have a number of $s$ quarks 
slightly lower than the number
of $u$ or $d$ quarks \cite{Kasuya}. Thus the nuclearite core should have a 
positive electric charge which
would be balanced by a number of electrons 
$N_e \simeq (N_d-N_s)/3$, where $N_d,~N_s$
and $N_e$ are the numbers of quarks $d,~s$ and 
electrons, respectively,  assuming $N_d=N_u$ \cite{Kasuya}.  \par

\begin{figure}
\vspace{-1.3cm}
\begin{center}
\mbox{  \epsfysize=9.5cm 
        \epsffile{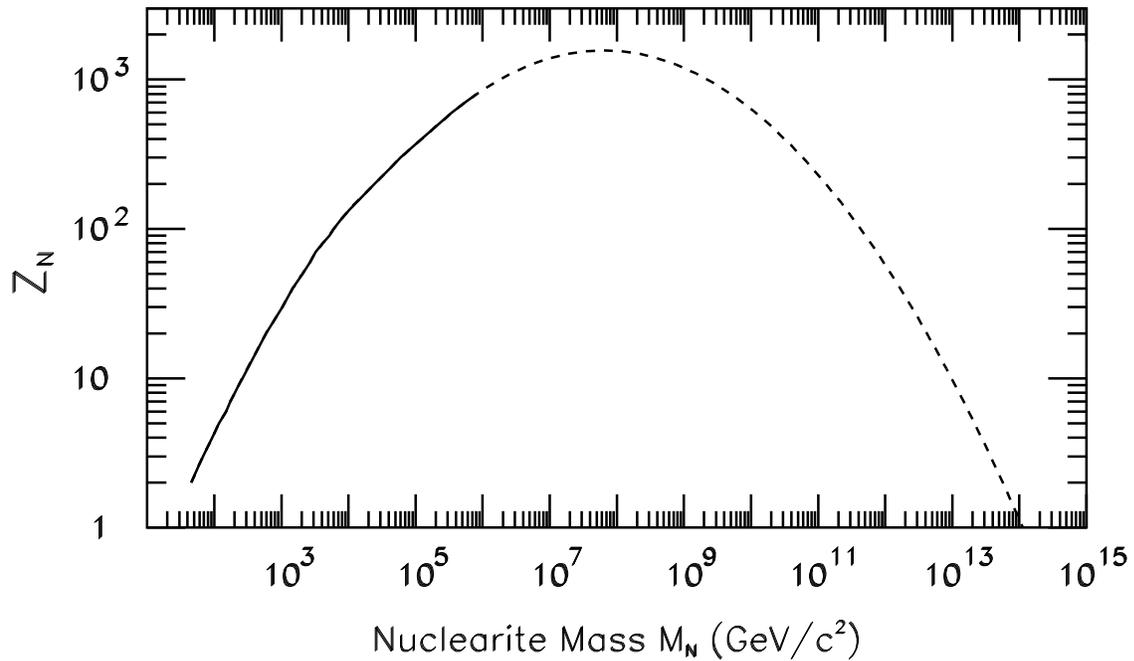}}
\end{center}
\vspace{-1cm}
\caption{Qualitative dependence of the nuclearite core charge $Z_N$ 
from the nuclearite 
mass  $M_N$. The solid line is computed in ref. [13];  the dashed line is a
rough interpolation which
takes into account that for $M_N \geq 10^{15}$ GeV/c$^2$ all the electrons are 
inside  the quark bag.}  
\label{fig:a-z}
\end{figure}

In the following, the radii $R_N$ refer only to the nuclearite
core (strangelet); the core+electronic cloud (nuclearite) system
should have the constant radius of $\sim 1$ \AA~for $R_N< 1$ \AA~ $=10^5$ 
fm. \par
For $R_N \geq 10^{5}$ fm,
 all electrons must be inside the quark bag. \par
For $10^{4} < R_N< 10^{5}$ fm, a fraction of the electrons 
are inside the 
quark bag, another fraction is external and gives the global
dimension of $\sim
10^{-8}$ cm $= 10^{5}$ fm to the nuclearite 
core + electrons system. In this condition a nuclearite is similar to 
a Bohr atom. \par 
 Fig. \ref{fig:nucleariti} illustrates qualitatively the space distributions
for the system of nuclearite core+electrons. Notice that for $R_N> 10^{5}$
fm, we picture the nuclearite as a sort of Thomson atom. \par
 For nuclearite
masses larger than $1.5 \cdot 10^{-9}$ g $\simeq 10^{15}$ GeV/c$^2$, the 
relation between mass and radius
should be (mass $\propto$ volume, $M_N \propto V_N \propto R_N^3$) 
\begin{equation}
R_N= \left( \frac {3M_N}{4 \pi \rho_N} \right)^{1/3} \label{eq:1}
\end{equation}

For $M_N=1.5 \cdot 10^{-9}$ g, the nuclearite should have a radius 
{\normalsize
\begin{equation}
R_N= \left( \frac {3}{4\pi} \frac{M_N}{3.5 \cdot 10^{14}} 
\right) ^{1/3} = 8.8  \cdot 10^{-6}~M_N^{1/3}=8.8  \cdot 10^{-6} \cdot 
(1.5 \cdot 10^{-9})^{1/3} \simeq 10^{-8}~\mbox{cm}=10^{5}~\mbox{fm}=
1~\mbox{\AA} 
\label{eq:2}
\end{equation}
}
For $M_N=10^{18}$ GeV/c$^2$ one has 
$R_N \simeq 10^5$ fm $(1000)^{1/3}=10^6$ fm. \par
Assuming that formula (\ref{eq:1}) is valid also for $M_N< 10^{15}$ 
GeV/c$^2$, one has, for the nuclearite core, the following values \par 

$$M_N=10^{18}~\mbox{GeV/c}^2 \hspace{3cm}  R_N\simeq 10^{6}~ \mbox{fm}$$   
$$ \hspace{1.2cm} M_N=10^{15}~\mbox{GeV/c}^2 \hspace{3cm}  R_N \simeq 10^{5} 
~\mbox{fm}=1~\mbox{\AA} $$
$$M_N=10^{9}~\mbox{GeV/c}^2 \hspace{3cm}  R_N \simeq 10^{3} ~\mbox{fm} $$
$$M_N=10^{3}~\mbox{GeV/c}^2 \hspace{3cm}  R_N \simeq 10~ \mbox{fm} $$

 Fig. \ref{fig:r-vs-mass-nuc}
gives the dependence of the core radius $R_N$ and of the radius of the
core+electrons system on the nuclearite mass $M_N$. \par
The expected relation between mass $M_N$ and charge $Z_N$ 
for nuggets of strange quark matter 
is shown in Fig. \ref{fig:a-z}. The solid line is the curve computed in ref.
\cite{Kasuya}, the dashed line is a rough interpolation taking into 
account that for $M_N \geq 10^{15}$ GeV/c$^2$ all the electrons are 
inside the quark bag. Nuclearites with $\beta \sim 10^{-3}$ could be
detected by scintillators and nuclear track detectors, independently 
of the charge of the nuclearite core. \par
A ``curious" problem arose in the discussion of the nuclearite production
in heavy ion high energy colliders: there was a fear that possibly produced
SQM would grow in size and destroy the earth. This was
 proven to be inaccurate in ref. \cite{Dar}. \par
The present flux upper limits on nuclearites in the cosmic radiation
are discussed in Section
\ref{par:Darkmatter} and given in ref. \cite{Lim-nucl}. \par

\section{Q-balls}
\label{par:Q-balls}
Q-balls \cite{Coleman} are aggregates of 
squarks $\tilde q$, sleptons $~\tilde l$ and 
Higgs fields \cite{Kusenko1}.  The scalar condensate inside a Q-ball core
has a global baryon number $Q$ (may be also lepton number) 
and a specific energy much smaller 
than 1 GeV per
baryon. We assume that the $Q$ numbers of quarks and squarks are equal
to 1/3 ($Q_q=Q_{\tilde q}=1/3$) or 2/3 
($Q_q=Q_{\tilde q}=2/3$). Protons, neutrons and may be electrons
can be absorbed in the condensate. 

\begin{figure}[ht]
\begin{center}
\mbox{
        \epsfig{file=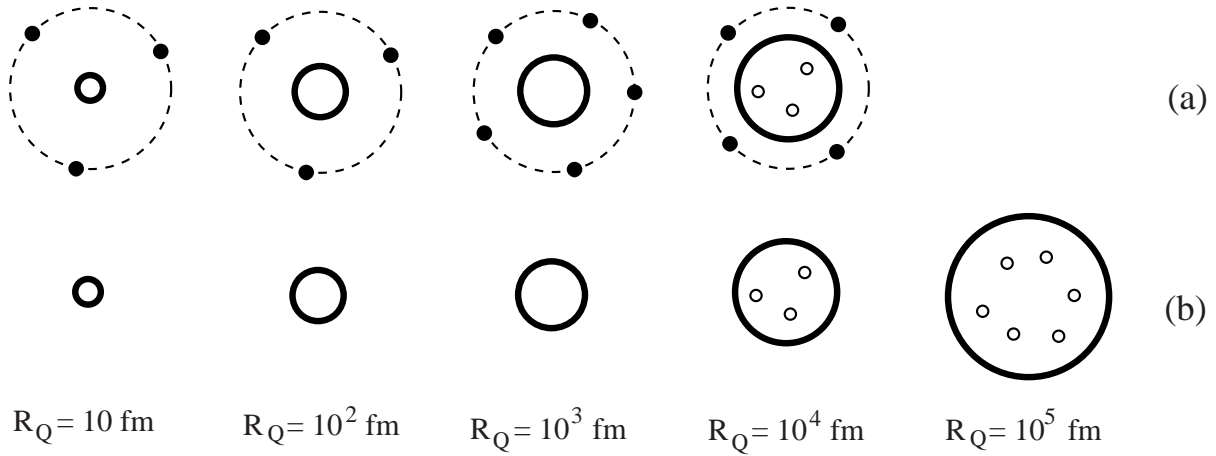,height=6cm}}
\end{center}
\vspace{-0.5cm}
\caption{Possible structure of  Q-balls: (a) SECS (charged core) and (b) SENS 
(neutral core). $R_Q$ (in a logarithmic scale) 
indicates the dimension of the 
squark condensate, the Q-ball core; the black points indicate 
electrons, open circles 
indicate s-electrons ($\tilde e$). For $R_Q > 10^4$ fm, the core contains 
$\tilde e$ (not $e^-$) and thus one should have only neutral Q-balls 
(SENS). $R_Q=10^6$ fm should be an upper limit for SENS radii.}
\label{fig:qballs}
\end{figure}

The vacuum expectation value inside a Q-ball core 
develops along  ``flat directions"
of the potential \cite{flat}. These flat directions are 
parametrized by combinations
of squarks and sleptons that are electrically neutral (otherwise they would
not be flat directions). Supposing that the different flavour squarks are not
mass degenerate, their numbers inside the Q-ball core would not be equal
for the same reason as in the nuclearite core case. \par
By assuming that the baryon number is packed inside 
 a Q-ball core, one can get upper limits for the Q-ball quantum number $Q$ and
for the Q-ball mass $M_Q$: $Q \leq 10^{30}$ and $M_Q \leq 10^{25}$ 
GeV/c$^2$, respectively; 
 Q-balls with $M_Q < 10^{8}$ GeV/c$^2$ are unstable \cite{Kusenko2}. \par
In the early universe only neutral Q-balls were produced:
 SENS (Supersymmetric Electrically Neutral Solitons), which
do not have a net electric charge, are generally massive and may 
catalyse proton decay. SENS may obtain an integer positive electric charge
absorbing a proton in their interactions with matter; thus we may have 
 SECS (Supersymmetric Electrically Charged Solitons), which
have a core 
electric charge,  have generally lower masses and the Coulomb
barrier prevents the capture of nuclei. SECS have only integer
charges because they are colour singlets. Some
Q-balls which have sleptons in the condensate can  also absorb
 electrons. The squarks $\tilde q$ inside the scalar 
potential bag have essentially zero mass. \par
The possible structures of 
SECS and SENS are shown in Fig. \ref{fig:qballs} a,b. \par
SENS may also interact with a proton of the 
interstellar medium,  catalyse the proton decay leading 
to the emission of $2-3$ pions (or kaons) and
transform quarks into squarks via the reaction $qq \to \tilde q \tilde q$. Thus
some SENS may become SECS with a charge $+1$ emitting $2~ \pi^0$ with total
energy of about 1 GeV.  \par
When a SENS enters the earth atmosphere, it could absorb, for example, a
nucleus of nitrogen which gives it the positive charge of $+7$ (SECS with
$Z= +7$). The next
nucleus absorption is prevented by Coulomb repulsion. If the  
Q-ball can absorb electrons at the same rate as protons, the positive charge
of the absorbed nucleus may be neutralized by the charge of absorbed 
electrons. The incoming SENS remain neutral most of the
 time. Electrons may be absorbed via the reaction $u+e \to d+ \nu_e$.
 If, instead, the absorption of electrons is slow or 
impossible, the Q-ball carries a positive electric charge after the capture
of the first nucleus in the atmosphere (SECS). 
 Other SENS could ``swallow" entire
atoms (remaining SENS). If a SENS could absorb an electron, it would acquire a 
negative charge (SECS with $Z=-1$). In the following we shall neglect
the possibility that a neutral Q-ball (SENS) becomes charged (SECS) 
and viceversa.

\begin{figure}[ht]
\begin{center}
\mbox{  \epsfysize=9cm
        \epsffile{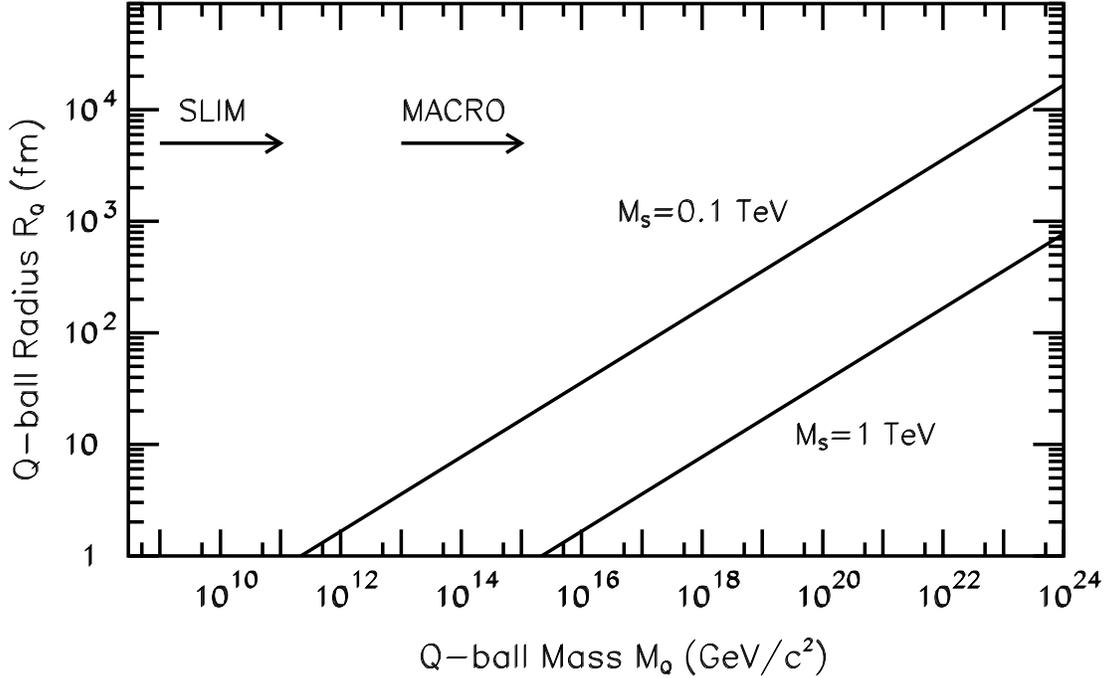}}
\end{center}
\vspace{-0.5cm}
\caption{Dependence of the Q-ball radius $R_Q$ from its mass $M_Q$ for two
values of the $M_S$ parameter. Q-balls with $M_Q \leq 10^8$ GeV/c$^2$ are 
unstable; Q-balls with $M_Q \geq 10^{25}$ GeV/c$^2$ should be very rare. We
also indicate the mass regions accessible by the SLIM and MACRO experiments
for Q-balls with $\beta \sim 10^{-3}$ in the cosmic radiation.}
\label{fig:qballsr-mass}
\end{figure}

The Q-ball mass $M_Q$, core size $R_Q$ and global quantum number $Q$ 
are related by the following relations \cite{Kusenko2}, in theories with a
bag potential $V=\phi(M_S^4)$,

\begin{equation}
M_Q=\frac{4 \pi \sqrt{2}}{3} M_S Q^{3/4} \simeq 
5924~M_S(\mbox{TeV})~Q^{3/4}~(\mbox{GeV})
\label{eq:1-qball}
\end{equation}
\begin{equation}
R_Q=\frac{1}{\sqrt{2}} M_S^{-1} Q^{1/4} \simeq 1.4 \cdot 10^{-17} ~M_S^{-1}
(\mbox{GeV}^{-1})~Q^{1/4}~\mbox{(cm)}
\label{eq:2-qball}
\end{equation}
where the parameter $M_S$ is the energy scale of the SUSY breaking 
symmetry. In the following we shall use $M_S=100$ GeV/c$^2$ and 
$M_S=1000$ GeV/c$^2$.\par
From  Eq. (\ref{eq:1-qball}) we have
\begin{equation}
Q^{1/4}=\left( \frac{3}{4 \pi \sqrt{2}} \frac{M_Q}{M_S} \right)^{1/3}
\label{eq:3-qball}
\end{equation}

Placing Eq. (\ref{eq:3-qball}) into Eq. (\ref{eq:2-qball}), we have
\begin{equation}
R_Q=\frac{1}{\sqrt{2}}M_S^{-4/3} \left( \frac{3M_Q}{4 \pi \sqrt{2}} 
\right)^{1/3} \label{eq:4-qball}
\end{equation}
Note that, as for nuclearites, we have $R_Q \sim M^{1/3}$, once $M_S$ 
is fixed. \par 
For $M_S=100$ GeV/c$^2$, we have from Eq. (\ref{eq:4-qball}):
$$ R_Q=\frac{1}{\sqrt{2}} \left( \frac{3M_Q}{4 \pi \sqrt{2}} \right)^{1/3}
\eqno (7\mbox{a}) $$
$$ R_Q(\mbox{fm}) \simeq 0.39~ M_Q(\mbox{GeV/c}^2)^{1/3} = 0.39~ 
(M_Q \cdot 0.197)^{1/3}~\mbox{fm}=0.227~M_Q^{1/3}~\mbox{fm} 
\eqno(7\mbox{b})$$
The dependence of $R_Q$ from $M_Q$ is shown in 
Fig. \ref{fig:qballsr-mass} for $M_S=100$ GeV/c$^2$ and 1000 GeV/c$^2$. \par
\setcounter{equation}{7}

The Q-balls have been considered as possible cold dark matter 
candidates; their core sizes should be only one order of
magnitude larger than a typical atomic 
nucleus \cite{Kusenko2,Kusenko2b}.\par
Flux limits on Q-balls come
mainly from the astrophysical dark matter limit given in 
Fig.~\ref{fig:Darkmatter}. SECS
with $\beta \simeq 10^{-3}$ and $M_Q < 10^{13}$ GeV/c$^2$ could reach 
an underground detector from above, SENS also from below 
\cite{Kusenko2,Kusenko3}.
 SENS may be detected by their almost continuous emission of charged pions 
(energy loss of about 100 GeV g$^{-1}$cm$^{2}$), while SECS may be detected by
their large energy losses yielding light in scintillators, and 
possibly ionization. The energy losses of Q-balls in matter were 
computed in ref. \cite{loss-qballs}.
 Flux upper limits on SECS could be deduced from the limits for dyons
with the same electric charge; flux limits on SENS could be 
obtained from limits on MMs which catalyse proton decay.

\section{Astrophysical limits}
\label{par:Darkmatter}

Magnetic monopoles, nuclearites and Q-balls could be components of the
galactic cold dark matter, required by the rotation curves of the stars
in the outskirts of our galaxy and of other galaxies.
 Assuming a local DM energy density of $\rho=0.3$ GeV/cm$^3$ and that 
MMs, nuclearites and/or Q-balls could be part of it and have typical 
velocities of $\beta \simeq 10^{-3}$, we can obtain upper limits
on their flux in the cosmic radiation. \par

\begin{figure}[ht]
\vspace{-3.5cm}
\begin{center}
\mbox{\epsfysize=14.5cm
        \epsffile{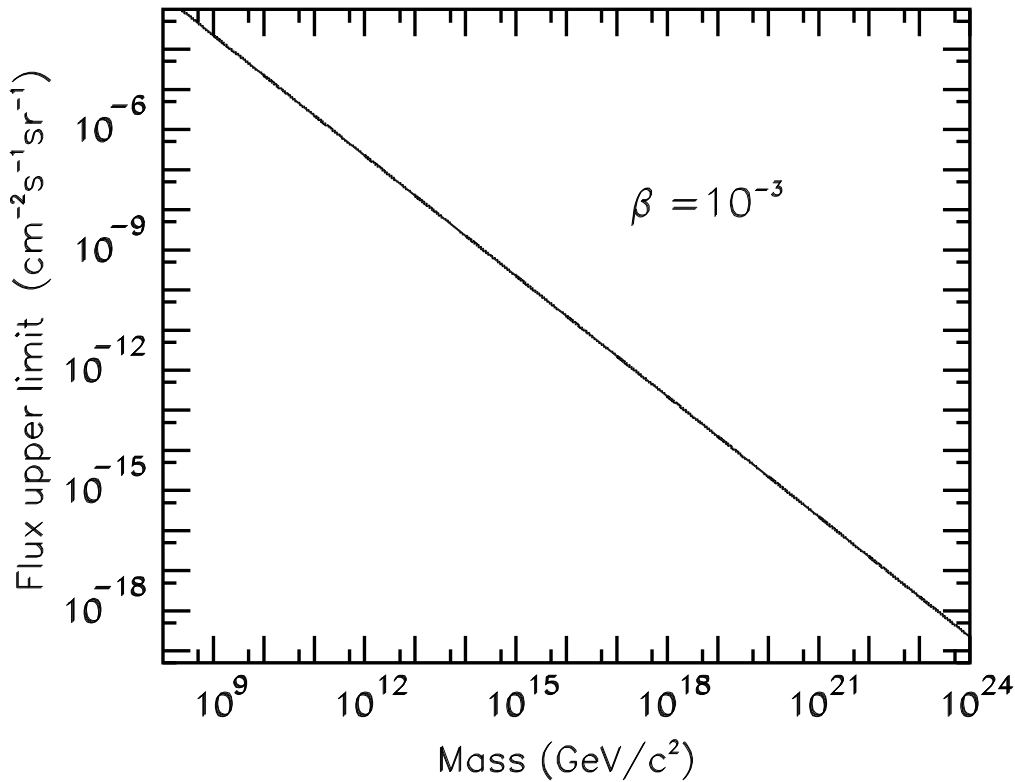}}
\end{center}
\vspace{-0.5cm}
\caption{Flux upper limits for nuclearites and for Q-balls 
 versus their masses, assuming that they have $\beta = 10^{-3}$ and 
that each of them saturates the local dark matter density. Clearly, if
the abundance of each of them is $10^{-3}$ of the cold dark 
matter, the quoted limits are $10^{-3}$ times smaller.}
\label{fig:Darkmatter}
\end{figure}

Fig. \ref{fig:Darkmatter} shows the flux upper limits for nuclearites 
and for Q-balls versus their mass, assuming that they saturate the local DM
density. Clearly, if their percentages are only $1\%$ of the local DM, then
the limits have to be lowered by a factor of 100. \par
An experiment like MACRO has placed upper limits on the fluxes of MMs
and nuclearites and could place similar upper limits for SECS at the level
of few $10^{-16}$ cm$^{-2}$s$^{-1}$sr$^{-1}$ for masses larger than 
approximately $10^{14}$
GeV/c$^2$ \cite{Monopoli,Lim-nucl}, see 
Fig. \ref{fig:qballsr-mass}. The experiment 
should also be capable to place limits on the MM catalysis of 
proton decay. \par
The SLIM experiment could reach a level of sensitivity 
 of few $10^{-15}$ cm$^{-2}$s$^{-1}$sr$^{-1}$ for masses larger
than approximately $10^{8}$ GeV/c$^2$ \cite{slim}.\par
For superheavy MMs the DM limits are considerably larger than the present
experimental limits. But monopoles can gain energy in the 
galactic field; from the survival of the galactic field one has more 
stringent limits at the level of $\sim 10^{-15}$ cm$^{-2}$s$^{-1}$sr$^{-1}$
(Parker limit \cite{Parker}), or even orders of magnitude lower
(Extended Parker limits \cite{Extended}). The limits, both 
experimental and expected, are less well known for intermediate mass
monopoles.

\section{Conclusions}
\label{par:Conclusions}
We have discussed qualitative pictures of the structures of superheavy 
and intermediate mass magnetic monopoles, of nuclearites and of Q-balls. In
particular we have given the possible structure of these objects as function
of their mass. We concluded with astrophysical considerations on their 
flux limits in the cosmic radiation. For nuclearites and Q-balls we assumed 
that they could be part of the cold DM and have $\beta = 10^{-3}$. Magnetic 
monopoles may be accelerated by galactic and intergalactic magnetic fields to
higher velocities; for MMs, DM limits are considerably higher than limits
based on the influence of magnetic fields.

\begin{center}
{\bf Acknowledgments}
\end{center}

We thank many members of the MACRO Collaboration for cooperation
 and for providing information and criticisism. We thank A. Kusenko for 
many stimulating 
discussions.

\end{document}